\documentstyle[preprint,prb,aps]{revtex}
\begin{document}

\draft
%\twocolumn[
%\widetext

\title{Compressible Strips, Chiral Luttinger Liquids, \\
and All That Jazz}
\author{A.H. MacDonald}
\address{Department of Physics, Indiana University, Bloomington,
IN 47405, USA.}

\date{\today}
\maketitle

\begin{abstract}
When the quantum Hall effect occurs in a two-dimensional electron gas, all
low-energy elementary excitations are localized near the system edge.  The
edge acts in many ways like a one-dimensional ring of electrons,  except
that a finite current flows around the ring in equilibrium. This article
is a brief and informal review of some of the physics of quantum Hall
system edges. We discuss the implications of macroscopic {\em compressible
and incompressible strip} models for microscopic {chiral Luttinger liquid}
models and make an important distinction between the  origin of
non-Fermi-liquid behavior in fractional quantum Hall edges and in usual
one-dimensional electron gas systems.
\end{abstract}


\section{Introduction}

The quantum Hall effect\cite{qhedisc} is an anomaly which occurs in the
transport properties of two-dimensional electron system in the regime of
strong perpendicular magnetic fields. At certain magnetic fields it is
found that the voltage drop in the system in the direction of current flow,
responsible for dissipation in the system,
vanishes at low temperatures. Our
understanding of this transport anomaly is not absolutely complete.
Nevertheless, there is fairly broad agreement that edge
state\cite{edgepicture,lhreview} theories, in which the nonequilibrium
transport currents are carried near the edge of the system, capture the
essence of the phenomenon. We present a version of the edge state
picture which accommodates both integer and fractional quantum Hall
effects in Section II of this article. It follows from the edge state
picture that when the quantum Hall effect occurs, the ground state of
the two-dimensional electron system {\em must} possess a set of low-energy
excitations which are localized near the edge of the system. In this
article we discuss a sequence of pictures of edge excitations in the
quantum Hall effect. The order in which we present the pictures is loosely
one of increasing sophistication and power, although there are important
lessons at each step and the more powerful descriptions are not in every
sense the ones with greater generality. In Section III we discuss edge
states for a system of non-interacting electrons in which the particles
are confined to a finite area by an external potential. In Section IV we
grapple with the Coulomb interaction between electrons. The
long-range of this interaction can have overwhelming importance in
determining the edge electronic structure when the edge charge density
profile has a width which is long compared to microscopic lengths, the
so-called `smooth edge' regime. The discussion of Section IV uses both
Thomas-Fermi and Hartree-Fock approximations to treat electron-electron
interactions. A many-particle description beyond the Hartree-Fock
approximation is necessary to describe edge states in the case of the
fractional quantum Hall effect. In Section V we discuss the ground state
and excited states of edges in the fractional case using a language of
many-body wavefunctions. This approach shows that the edges are very
similar in integer and fractional cases but fails to reveal some
important differences which appear in quantities like the tunneling
density-of-states at the edge of the system. We address these issues by
using Wen's\cite{wen,wenreview} chiral Luttinger liquid theory to discuss
integer and fractional edges in Section VI. In Section VII we present some
concluding remarks.

\section{Incompressibility and Edge States}

The thermodynamic compressibility of a system of interacting particles is
proportional to the derivative of the chemical potential with respect to
density. It can happen that at zero temperature the chemical potential
has a discontinuity at a density
$n^{*}$: the energy to add a particle to the system
($\mu^{+}$) differs, at this density, from the energy to remove a
particle from the system ($\mu^{-}$).  The system is then said to be
incompressible.  In an incompressible system a finite energy is
required to create independent positive and negative charges
which are capable of carrying current through the bulk. The number of
these free charges present in the system will have an activated
temperature dependence and will vanish for $T \to 0$. Incompressible
systems are usually insulating. Paradoxically, as we explain below,
incompressibility is precisely the condition required for the quantum
Hall effect to occur. The twist is that in the case of the quantum Hall
effect, the density $n^{*}$ at which the incompressibility occurs must
depend on magnetic field. In my view {\em incompressibility at a
magnetic-field-dependent density} is the key to the quantum Hall effect.
This property requires the existence of the gapless edge excitations
which are the subject of this article.

To make this more concrete, consider a large but finite two-dimensional
electron gas at zero temperature, as illustrated in
Fig.~[\ref{fig:incomp}].  We consider the case in which the chemical
potential lies in the `charge gap'; $ \mu \in (\mu^{-},\mu^{+})$. We want
to consider the change in the equilibrium local currents, present in the
system because of the breaking of time-reversal-invariance by the magnetic
field, when we make an infinitesimal change in the chemical potential,
$\delta \mu$. Because $\mu$ lies in the charge gap the change in the
local current density anywhere in the bulk of the system must be zero. The
current density can change, if it does anywhere, only at the edge of the
system. It follows from charge conservation that, if there is a change in
the current flowing along the edge of the system, it must be the same at
any point along the edge. We can relate this change in current to the
change in the orbital magnetization:
\begin{equation}
\delta I = {c \over A} \delta M.
\label{eq:orbmag}
\end{equation}
Eq.~(\ref{eq:orbmag}) is just the equation for the magnetic moment of a
current loop. However,
\begin{equation}
\delta M = {\partial M \over \partial \mu}\vert_B \, \delta \mu =
{\partial N \over \partial B}\vert_{\mu} \, \delta \mu.
\label{eq:maxwell}
\end{equation}
The second equality in Eq.~(\ref{eq:maxwell}) follows from a Maxwell
relation. Combining Eq.~(\ref{eq:orbmag}) and Eq.~(\ref{eq:maxwell}) we
obtain the following result for the rate at which the equilibrium edge
current changes with chemical potential when the chemical potential lies
in a charge gap:
\begin{equation}
{\delta I \over \delta \mu} = c {\partial n \over \partial B}|_{\mu}.
\label{eq:streda}
\end{equation}
The fact that $\delta I / \delta \mu \ne 0$ implies that whenever the
charge gap occurs at a density which depends on magnetic field, there {\em
must} be gapless excitations at the edge of the system. From a more
microscopic point of view, Eq.~(\ref{eq:streda}) arises because the edge
currents are related to the way in which the spectrum evolves with changes
in the the vector potential and hence in the magnetic
field.\cite{edgepicture}

This property of the edge states is expected to persist even if the
chemical potential lies only in a mobility gap and not in a true gap, as
illustrated schematically in Fig.~[\ref{fig:incomp}]. A net current can be
carried from source to drain across the system by changing the local
chemical potentials only at the edges and having different chemical
potentials along the two edges connecting source and drain.  When bulk
states are localized, the two edges and the bulk are
effectively decoupled from each other. Eq.~(\ref{eq:streda}) then
also applies to transport currents, relating the current carried
from source to drain to the chemical potential
difference between the two edges, equal to
$e V_H$ where $V_H$ is the Hall voltage.  There is no
voltage drop along an edge since each edge is in local equilibrium and
hence no dissipation inside the sample.
Eq.~(\ref{eq:streda}) was proposed as an explanation for
the quantum Hall effect by Pavel St\v{r}eda and
is commonly known as the St\v{r}eda formula.\cite{streda}

In using this picture to explain transport experiments in bulk systems it
is necessary to claim that the transport current will be carried entirely
at the edge of the system even when bulk states occur at the Fermi level,
as long as these states are localized. There are difficulties with this
argument as a complete explanation for all transport phenomena associated
with the quantum Hall effect, but that is another story and we will not
pursue it here. In our view, however, there is no difficulty with the
conclusion relevant to the present paper; gapless edge excitations
satisfying Eq.~(\ref{eq:streda}) must be present whenever the quantum
Hall effect occurs.

\section{Non-Interacting Electron Picture}

Throughout this article we will consider a disk geometry where electrons
are confined to a finite area centered on the origin by a circularly
symmetric confining potential,
$V_{\rm conf}(r)$. We have in mind the situation where $V_{\rm conf}(r)$ rises
from zero to a large value near $r = R$, where $R$ is loosely speaking the
radius of the disk in which the electron system is confined. We choose
this geometry, for which the electron system has a single edge, since we
limit our attention here to the properties of an isolated quantum Hall
edges and will not discuss the physics of interaction or scattering
between edges.\cite{twoedgecaveat} In this geometry it is convenient to
choose the symmetric gauge,
$ \vec A = B (-y,x,0)/2$. For $V_{\rm conf}(r) \equiv 0$, the single-electron
kinetic energy eigenfunctions and eigenenergies in this gauge have
definite angular momentum and are known analytically.\cite{lhreview} The
spectrum consists of a set of degenerate Landau levels. The kinetic
energy in the n-th Landau level is $\hbar \omega_c (n+1/2)$ where
$\omega_c = e B / m c$ is the cyclotron frequency; states occur in the
n-th Landau level with\cite{negative} angular momentum $m = -n, -n +1,
\cdots$. In each Landau level states with larger angular momentum are
localized further from the origin. For example, the symmetric gauge
wavefunctions in the lowest Landau level are:
\begin{equation}
\phi_m(z) =\big( \frac{ 1} {2 \pi \ell^2 2^m m!} \big)^{1/2}
z^m \exp ( - |z|^2 /4 \ell^2 )
\label{eq:phim}
\end{equation}
where $2 \pi \ell^2 B = h c/e $ is the magnetic flux quantum and $\ell$
is the magnetic length which serves as the fundamental microscopic length
scale in the strong magnetic field regime. It is easy to verify that, for
large
$m$ \ , $\phi_m$ is localized near a circle with radius
$R_m = \sqrt{2(m+1)} \ell$. (Note that for large $m$ the separation
between adjacent values of $R_m$ is $ \ell^2 / R_m << \ell$.)  In the strong
magnetic field limit, which we will adopt consistently in this article,
the confinement potential does not mix different Landau levels. Since
there is only one state with each angular momentum in each Landau level,
that means that the only effect of the confinement potential is to increase
the energy of the symmetric gauge eigenstates when $R_m$ becomes larger
than $\sim R$. The typical situation is illustrated schematically in
Fig.~[\ref{fig:nonintel}]. Here the $n=0$ and $n=1$ Landau levels are
occupied in the bulk and the chemical potential $\mu$ lies in the gap
$\Delta =
\hbar \omega_c$ between the highest energy occupied Landau level ($E = 3
\hbar \omega_c /2$) and the lowest energy unoccupied Landau level ($E = 5
\hbar \omega_c/2$). The states we are interested in in this article are
the ground state and the low energy excited states obtained by making one
or more particle-hole excitations at the edge. We will actually discuss
only the simplest situation where a single Landau level crosses the
chemical potential at the edge of the system and the analogous {\em
single branch} situations in the case of the fractional quantum Hall
effect.\cite{impatedge} We will also neglect the spin degree of freedom
of the electrons throughout this article.

An important property of the ground state of the non-interacting electron
system in the case of interest, is that it remains an exact eigenstate of
the system (but not necessarily the ground state!) when interactions are
present. That is because the total angular momentum $K$ for this state
is
\begin{equation}
M_0 = \sum_{m=0}^{N-1} m = N (N-1) /2
\label{nov1a}
\end{equation}
and all other states in the Hilbert space (truncated to the lowest Landau
level) have larger angular momentum.\cite{ajp} For large disks and total
angular momentum near $M_0$ the excitation energy of a non-interacting
electron state will be
\begin{equation}
\Delta E = \gamma M
\label{nov1aa}
\end{equation}
where $M \equiv K - M_0$ is the excess angular momentum and $\gamma$ is
the energy separation between single-particle states with adjacent angular
momenta and energies near the Fermi energy. $\gamma$ is related to the
electric field, $E_{\rm edge}$ from the confining potential at the edge of
the disk:
\begin{equation}
\gamma = e E_{\rm edge} \frac{d R_m}{d m}
 = e E_{\rm edge} \ell^2 / R
\label{nov1b}
\end{equation}
This expression for $\gamma$ can be understood in a more appealing way. In
a strong magnetic field charged particles execute rapid cyclotron orbits
centered on a point which slowly drifts in the direction perpendicular to
both the magnetic field and the local electric field. For an electron at
the edge of the disk the velocity of this `E cross B' drift is $v_{\rm edge}
= c E_{\rm edge}/ B $. The energy level separation can therefore be written in
the form
\begin{equation}
\gamma = \hbar v_{\rm edge}/ R_{\rm edge} = h / T
\label{nov1c}
\end{equation}
where $T$ is the period of the slow drift motion of edge electrons around
the disk, in agreement with expectations based on semiclassical quantization.

Since the excitation energy depends only on the angular momentum increase
compared to the ground state it is useful to classify states by $M$. It
is easy to count the number of distinct many-body states with a given
value of $M$ as illustrated in Fig.~[\ref{fig:fermioncounting}]. For
$M=1$ only one many-particle state is permitted by the Pauli exclusion
principle; it is obtained by promoting the ground state electron with
$m=N-1$ to $m=N$. For $M=2$, particle hole excitations are possible from
$m=N-1$ to
$m=N + 1$ and from $m=N-2$ to $m= N$. In general $M$ many-particle
states with excess angular momentum $M$ can be created by making a
single-particle hole excitation of the ground state. For $M \ge 4$
additional states can be created by making multiple particle-hole
excitations. The first of these is a state with two particle-hole
excitations which occurs at $M=4$ and is illustrated in
Fig.~[\ref{fig:fermioncounting}].

\section[]{Compressible and Incompressible Strips:\\ The Thomas-Fermi
Picture}

It is very instructive to apply a Thomas-Fermi approximation to the edge
of an electron disk in the quantum Hall regime.\cite{tfapprox} A modern
framework for discussing the Thomas-Fermi approximation is provided by
density-functional theory, which has been generalized in recent years to
accommodate magnetic fields.\cite{giovanni} The Thomas-Fermi approximation
is intended to be applicable to the case where the charge density changes
very slowly on atomic length scales. In this regime we can hope that the
local-density-approximation, where the energy density at $\vec r$
is assumed to be equal to the energy density of a uniform
system with density $n = n(\vec r)$, is valid.
The main lessons to be learned from
the Thomas-Fermi approximation for quantum Hall edges are connected with
the long-range of the Coulomb interaction and our discussion here is
always for the case of these physically realistic interactions. When the
Thomas-Fermi approximation is applied for systems of charged particles, the
electrostatic interaction energy, for which a local density approximation
is obviously inappropriate, must be treated separately. The Thomas-Fermi
energy functional is therefore
\begin{equation}
E[n] = E_H[n] + \int d^2 \vec r n (\vec r) \epsilon (n(\vec r) )
+ \int d^2 \vec r n(\vec r) V_{\rm ext} (\vec r),
\label{nov2a}
\end{equation}
where $E_H[n]$ is the electrostatic (Hartree) energy, $V_{\rm ext}$ is the
external potential, and $\epsilon (n)$ is the energy per-particle of a
uniform density system (in a magnetic field in our case) which is placed
in an electrically neutralizing positively charged background to
eliminate the electrostatic energy. Minimizing this energy functional
with respect to $n (\vec r)$ at fixed total particle number gives the
Thomas-Fermi approximation expression for the density profile,
\begin{equation}
V_{\rm ext} (\vec r) + V_H (\vec r) + \mu_{2D} (n (\vec r)) = \mu.
\label{nov2b}
\end{equation}
Here $V_H(\vec r)$ is the (non-local) electrostatic potential,
$\mu$ is the chemical potential, and
\begin{equation}
\mu_{2D}(n) = \frac{d \, n \epsilon(n)}{d n}
\label{nov2c}
\end{equation}
is the density-dependent chemical potential of a uniform density electron
system in a neutralizing background.

Eq.~(\ref{nov2c}) can and has been used to discuss the density profile at
the edge in the case of both integer\cite{tfapprox} and
fractional\cite{ferconi} quantum Hall effects. We will discuss here
the Thomas-Fermi version of the random-phase-approximation, where we
include only the electrostatic and kinetic energy contributions to the
energy and fractional quantum Hall effect features are not captured. In
this approximation (neglecting spin for simplicity)
\begin{equation}
\mu_{2D}(n) = \hbar \omega_c ([\nu] +1/2)
\label{nov6a}
\end{equation}
where $[\nu]$ is the integer part of the Landau level filling factor $\nu
= 2 \pi \ell^2 N/A$. Let's assume for the moment that the non-interacting
electron ground state remains the ground state in the presence of
interactions. Looking on the macroscopic length scale appropriate to the
Thomas-Fermi approximation, the charge density of this state is constant:
$n = (2 \pi \ell^2 )^{-1}$ for $ R < R_{N}$ and $n = 0$ for $R > R_{N}$.
The electrostatic potential produced by this charge density is
\begin{equation}
V_H(r) = \frac{e^2}{\epsilon \ell} \sqrt{2N} F({1 \over 2},{-1 \over
2},1; {r^2 \over R_N^2}).
\label{nov6b}
\end{equation}
where $F(a,b,c;x)$ is the confluent hypergeometric function.

The important point for us is that the scale of this potential is larger
than the microscopic interaction energy scale
$ e^2/\ell $ by a factor proportional to $\sqrt {N}$. Since
$e^2/\ell$ and $\hbar \omega_c$ are of the same order it follows that
except for the case of small $N$ `quantum dot'
systems,\cite{qdot,qdotexp,qdottheory,largem} the maximum density droplet
wavefunction cannot be the ground state {\em unless} the sum of an
external potential and the Hartree electrostatic potential is close to
constant, {\em i.e.} unless the external potential is similar to that
from a neutralizing positive background which is co-planar with the
electrons. When this is not the case, the charge density profile will be
close to that which would be obtained in a purely electrostatic theory by
solving the equation
$V_{\rm ext}(r) + V_H(r) = \mu$. When $\mu_{2D}(n)$ is included, the solution
of the Thomas-Fermi equation for the charge-density will consist of
relatively broad regions where
$V_{\rm ext}(r) + V_H(r) = \mu - \hbar \omega_c (N +1/2)$ for some fixed
integer
$N$ and the charge density varies, and relatively narrow regions
separating adjacent values of $N$ where the charge density is fixed.

In this picture of quantum Hall system edges, regions where the electron
density varies are referred to as compressible strips and regions where
the electron density is fixed are referred to as incompressible strips.
Evidently, low-energy excitations can be created in the compressible
strips by varying the charge density along the edge or by altering
slightly the charge density profile across the edge. To relate the
Thomas-Fermi theory results to microscopic theory, we must, at a minimum,
use a Hartree or Hartree-Fock approximation. The results of such a theory are
illustrated schematically in Fig.~[\ref{fig:thfehf}]. In this
illustration we have in mind a `smooth edge' where the external potential
is created by charges on gates or surface states some distance from the
two-dimensional electron system. The extreme limit of the `smooth edge'
model systems is one where the external potential is taken to be
parabolic. For this parabolic confinement model, which is often
applicable to quantum dot systems, the electrostatic approximation to the
Thomas-Fermi equations can be solved
analytically:\cite{karenrefs,tfapprox}
\begin{equation}
n(r) = n_0 \sqrt{1 - (r/R)^2},
\label{eq:nov9}
\end{equation}
where $n_0$ and $R$, the radius of the electron disk, are constants fixed
by the total electron number and the curvature of the parabolic
potential. When $\mu_{2D}(n)$ is included in the Thomas-Fermi equations,
the density will distort slightly and develop the compressible and
incompressible strips discussed above and illustrated in
Fig.~[\ref{fig:thfehf}].
For smooth edges, electrostatic considerations will be so dominant that
nearly the same density profile will be produced by any microscopic
theory. The simplest microscopic theory that includes interactions is a
Hartree\cite{bpt} or Hartree-Fock theory in which the energy of
each single particle state is
modified by electrostatic (or Hartree) and, in the Hartree-Fock
case, exchange corrections. In such
a theory the regions in space where the local density does not correspond
to an integral Landau level filling factor, {\em i.e. the compressible
strips} must have fractional occupation numbers for the Hartree-Fock
single-particle states. It follows that the total variation of the
Hartree-Fock energy across the angular momenta in each compressible strip
must be smaller than the thermal energy $k_B T$. One point of view on
this result is to regard the system as locally metallic in the
compressible strips so that the confinement potential is strongly
screened.\cite{gerhards} Of course, the Hartree-Fock approximation is not
justified in this regime but it does provide one valid lesson. Throughout
the compressible region, single-particle states with nearby angular
momenta will have a finite probability of being occupied or
unoccupied as the result of thermal
or, in a more general theory, quantum fluctuations.  The sharp Fermi
edge between occupied and unoccupied states that we would have for
non-interacting electrons is lost.
It seems clear that the spectrum of low-energy
excitations can be exceedingly complicated in the smooth edge regime.

In the following sections we will implicitly assume that the external
confining potential permits the electron density at the edge to fall off
on a microscopic length scale, {\em i.e.} that we are in the `abrupt
edge' regime. We will also implicitly assume that the interactions between
the particles are short-ranged, appealing if pressed to the ubiquitous
presence of nearby gates which dress all electrons with image charges.
You have been fairly warned that these assumptions can be dangerous
especially when electrostatic imperatives force an electron-density that
changes slowly on microscopic length scales. The task of determining the
excitation energy (temperature), if any, below which the `abrupt edge'
models we will now discuss apply for `smooth edge' systems remains an
important challenge. In quantum dot systems the transition between
`abrupt edge' and `smooth edge' regimes is initiated by {\em edge
reconstructions}.\cite{qdottheory} Similar `reconstructions' may occur
at the edges of bulk systems and when they occur they will complicate the
excitation spectrum and all physical properties.

\section{Many-Body Wavefunction Picture}

In this section we discuss the edge excitation spectrum of interacting
electrons using a language of many-particle wavefunctions. For the case
of the integer quantum Hall effect we will essentially recover the
picture of the excitation spectrum obtained previously for non-interacting
electrons by counting occupation numbers. We could have used the
Hartree-Fock approximation and occupation number counting to generalize
these results to interacting electrons. However, the Hartree-Fock
approximation is completely at sea when it comes to the fractional case.
Discussions of the fractional edge using an independent electron language
are can be comforting but are, in my view,
misleading. Nevertheless, we will see that there is a one-to-one
correspondence between the edge excitation spectrum for non-interacting
electrons at integer filling factors and the fractional edge excitation
spectrum.

Many-electron wavefunctions where all electrons are confined to the
lowest Landau level must be sums of products of one-particle wavefunctions
from the lowest Landau level. From Eq.~(\ref{eq:phim}) it follows that
any $N$ electron wavefunction has the form
\begin{equation}
\Psi[z] = P(z_{1},\ldots ,z_{N})\; \prod_{\ell} \exp{(-|z_{\ell}|^{2}/4)},
\label{eq:nov10a}
\end{equation}
where we have adopted $\ell$ as the unit of length and
$P(z_1,\ldots,z_N)$ is a polynomial in the two-dimensional complex
coordinates. This property\cite{analytic} of the wavefunctions will be
exploited in this section. The first important observation is that since
$\Psi[z]$ is a wavefunction for many identical fermions it must change
sign when any two particles are interchanged, and therefore must vanish
as any two particles positions approach each other. Since
$P(z_1,\ldots,z_N)$ is a polynomial in each complex coordinate it
follows\cite{murray} that
\begin{equation}
P(z_{1},\ldots ,z_{N}) = \prod_{i<j} (z_{i} - z_{j})\; Q[z]
\label{eq:nov10b}
\end{equation}
where $Q[z]$ is any polynomial which is symmetric
under particle interchange. It is important to note that the total
angular momentum of all the particles ($K$) is just the degree of the
polynomial $P[z]$, {\em i.e.} the sum of the powers to which the
individual particle complex coordinates are raised. Since the total
angular momentum is a good quantum number the polynomial part of any
many-electron eigenstate will be a homogeneous degree polynomial, {\em
i.e.} all terms in the many-particle polynomial must have the same degree.
Additionally, the total angular momenta corresponding to a polynomial
which, as in Eq.~(\ref{eq:nov10b}), is the product of two polynomials is
the sum of the angular momenta associated with those polynomials.

\,From the discussion of Section III it is clear that for non-interacting
electrons in any monotonically increasing confinement potential, the
lowest energy state will be the state with the minimum total angular
momentum. In Eq.~(\ref{eq:nov10b}) that corresponds to choosing $Q[z]$ to
have degree zero, {\em i.e.} to $Q[z] \propto 1$. It is easy to verify
that the wavefunction when $Q[z]$ is a constant is in fact the Slater
determinant formed by occupying the single-particle states with $m=0,
\ldots, N-1$. For interacting electrons this state will remain the ground
state provided the confinement potential is strong enough to overcome the
repulsive interactions between electrons which favor states with larger
total angular momentum. When this is the ground state, low-energy
excited states with excess angular momentum
$M = K - N (N-1)/2$ are linear combinations of the states constructed by
choosing all possible symmetric polynomials\cite{stone} of degree $M$ for
$Q[z]$.

We'll discuss the enumeration of these polynomials in a moment
but pause now to explain how this analysis many be generalized to the
case of the fractional quantum Hall effect. We limit our attention
here\cite{macdedge} to the simplest fractional quantum Hall effects
which occur at Landau level filling factors $\nu =1/m$ for any odd
integer $m$; in some senses the $m=1$ case can be regarded as a special
case of the fractional quantum Hall effect. The physics of the chemical
potential jump which occurs at these filling factors was explained in the
pioneering paper of Laughlin.\cite{laughlin,lhreview} For $\nu < 1/3$, for
example, it is possible to find states in the Hilbert space in which
pairs of electrons are never found in a state with relative angular
momentum equal to one. This is the two-body state in which two electrons are
closest together.  All many-particle states which avoid placing pairs in
this state will have low energy.
If this condition\cite{caveat} is satisfied,
\begin{equation}
P(z_{1},\ldots ,z_{N}) \equiv \prod_{i<j} (z_{i} - z_{j})^{3}\; Q[z]
\label{eq:nov10c}
\end{equation}
for any symmetric polynomial $Q[z]$. If the system has an abrupt edge
the ground state will have $Q[z] \equiv 1$ just as in the non-interacting
case. The edge excitations correspond to the same set of symmetric
polynomials as in the $\nu =1$ case.  In the case of a model system with a
short-range interaction and a parabolic confinement potential,
it is easy to place\cite{largem} the argument we have sketched above on firm
ground.  It is known from numerical studies that the bulk chemical
potential discontinuity survives when the model interaction is
replaced by the realistic Coulomb interaction.  As long as the
external potential is such that we are in the `abrupt edge' regime,
the edge states in the realistic case will be in one-to-one correspondence with
those of Eq.~(\ref{eq:nov10c}).

The wavefunction
\begin{equation}
\Psi[z] \equiv Q[z] \prod_l \exp (- |z_l|^2/4)
\label{eq:nov10d}
\end{equation}
is a wavefunction for $N$ bosons in a strong magnetic field. Thus the
enumeration of the edge excitations in terms of symmetric polynomials
discussed above is equivalent to enumerating all many boson wavefunctions
with a given value of the total angular momentum. The boson angular
momentum
\begin{equation}
M = \sum_{m=0}^{\infty} m\; n_{m}
\end{equation}
where $n_m$ are the boson occupation numbers, is equivalent (for $\nu =
1/m$) to the excess angular momentum $M = K - m N (N-1)/2$ of the fermion
wavefunctions. In the state with $M=0$,
$n_0=N$, all other boson occupation numbers are zero, and
$Q[z]$ is a constant. In the boson language the ground state is a Bose
condensate. The lone state with $M=1$ has $n_1=1$,$n_0=N-1$; the
symmetric polynomial for this boson wavefunction is
\begin{equation}
Q[z] = z_{1} + z_{2} + \ldots z_{N}.
\label{eq:nov10e}
\end{equation}
For the integer $\nu =1$ case, it can be shown explicitly that the
corresponding many-fermion state is the $M=1$ state with a single
particle-hole excitation at the edge, discussed in Section III. The set
of excitations at general values of $M$ can be described equally well in
either fermion or boson languages. Some of the states which occur at
small values of $M$ are listed in Table I.

In the parabolic confinement case the total energy depends only on the
excess total angular momentum: $\delta E = \gamma M$. The number of
many-boson states with total angular momentum
$M$, $g(M)$ can be calculated by considering a system of non-interacting
bosons with single-particle energy $\gamma m$ so that $E = \sum_{m}
(\gamma m) \cdot n_{m} = \gamma M$. The partition function is
\begin{equation}
Z = \sum_{M} g(M)\; e^{-\gamma M/k_{B}T} = \sum_{M} x^{M}\; g(M)
\label{eq:nov12a}
\end{equation}
where $x = e^{-\gamma /k_{B}T}$. For $N \to \infty$, the $m=0$ state acts
like a reservoir with chemical potential $\mu = 0$ so that the partition
function calculation can be done in the grand canonical ensemble. The
degeneracies $g(M)$ can be read off the power series expansion of the
partition function:
\begin{eqnarray}
\lefteqn{Z = \prod_{k=1}^{\infty} \frac{1}{1 - x^{k}} = (1 - x)^{-1} (1
- x^{2})^{-1} (1 - x^{3})^{-1} \ldots}\nonumber\\
 &=& (1 + x + x^{2} + x^{3} + \ldots ) (1 + x^{2} + x^{4} + \ldots ) (1 + x^{3}
+ x^{6} + \ldots ) \ldots \nonumber\\
 &=& (1 + x + 2 x^2 + 3 x^3 + 5 x^4 + 7 x^5 + 11 x^6 +
 15 x^7 + 22 x^8 + 30 x^9 + 42 x^{10} + \ldots
 \label{eq:nov12b}
\end{eqnarray}
For\cite{largem} large $M$ $g(M) \sim e^{\sqrt{\frac{2}{3}}\cdot\pi\cdot
M^{1/2}}$. The function $g(M)$ is well known to number theorists from the
theory of partitions\cite{partitions} in which it is known as the
partition function, not to be confused with the physics partition
function above! For
parabolic confinement potentials and short-ranged repulsive
interactions, the degeneracy of the edge excitations at a given excess
angular momentum is exact in both integer and fractional $\nu =1/m$
cases. For general confinement potentials and general electron-electron
interactions these degeneracies will be lifted. However, there is reason
to expect that in the thermodynamic limit excitations with $M \ll
N^{1/2}$ will be nearly degenerate. One way to see this is to use the
chiral Luttinger liquid picture of quantum Hall edges which we discuss in
the following section. This approach will allow us to do more than
enumerate excitations of the system and, in particular will enable us to
discuss the density-of-states for tunneling into the edge of a quantum
Hall system.

\section{Chiral Luttinger Liquid Picture}

The chiral Luttinger liquid picture\cite{wenreview} of quantum Hall
systems is an adaptation of the Luttinger liquid theory of
one-dimensional electron systems. We start this section with a brief
outline of the portion of that theory that we require. Readers in search
of greater depth should look elsewhere.\cite{mahanll} As in higher
dimensions, low excitation energies states in a one-dimensional fermion
system will involve only single-particle states near the Fermi
wavevector. Since the differences in wavevector among the relevant
states at a given Fermi edge are small, the excitations produced by
rearranging them occur on length scales which are long compared to
microscopic lengths. It is therefore reasonable to argue that the energy
density in the system at any point in space should depend only on the
local density of left-moving ($k < 0$) and right-moving ($k>0$)
electrons, $n_L(x)$ and $n_R(x)$:
\begin{equation}
E[n_{L},n_{R}] = E_{0} + \int dx\;
\left[\frac{\alpha_{LL}}{2}\; \delta n_{L}^{2}(x) +
\frac{\alpha_{RR}}{2}\; \delta n_{R}^{2}(x) + \alpha_{LR}\; \delta
n_{L}(x)\; \delta n_{R}(x)\right].
\label{eq:enlnr}
\end{equation}
It is, perhaps, not completely obvious that the density provides a
complete parameterization of the low-energy excitations, and
indeed in the fractional Hall case there are situations where the analog
of Eq.~(\ref{eq:enlnr}) is incorrect.
Here $\alpha_{LL}$, $\alpha_{LR}$ and $\alpha_{RR}$ are determined by the
second derivatives of the energy per unit length with respect to $n_L$ and
$n_R$ for a uniform system and can be determined in principle by a
microscopic calculation.
$\delta n_L(x)$ and $\delta n_R(x)$ are differences of the density from
the ground state density. Note that we have as a convenience chosen the
chemical potential to be zero in dropping a term proportional to $ \int
dx (\delta n_L(x) + \delta n_R(x) ) $. We start by considering the case
where $\alpha_{LR} =0$ so that the left-moving electrons and right moving
electrons are decoupled. Focus for this case on the energy of the right
moving electrons. We Fourier expand the density and note that
\begin{equation}
\int dx\; \delta n_{R}^{2}(x) = \frac{1}{L} \sum_{q\neq 0} n_{-qR}.
n_{qR}
\label{eq:nov12c}
\end{equation}
so that the energy can be written in the form
\begin{equation}
E_{R} = E_{0} + \frac{\alpha_{LL}}{2L} \sum_{q\neq 0}
n_{-qR} n_{qR}.
\end{equation}

The energy above can be used as an effective Hamiltonian for low-energy
long-wavelength excitations. The simplification at the heart of
the Luttinger liquid theory is the observation that when the Hilbert
space is truncated to include only low-energy, long-wavelength excitations
(in particular when the number of left-moving and right-moving electrons
is fixed) Fourier components of the charge density do not commute. For
example consider the second quantization expression for $n_{qR}$ in terms
of creation and annihilation operators with $k > 0$:
\begin{equation}
n_{qR} = \sum_{k>0} c_{k+q}^{\dagger} c_{k}^{\phantom{\dagger}}.
\end{equation}

An example of the dependence of the effect of products of these operators
on the order in which they act is more instructive than the actual
algebraic calculation of the commutators. Note for example that
\begin{equation}
n_{-qR}\; |\Psi_{0}\rangle = 0
\label{eq:nov12d}
\end{equation}
where $q > 0$ and $|\Psi_0\rangle$ is the state with all right-going
electron states with $k < k_F$ occupied and all right-going states with $k
> k_F$ empty. (The alert reader will have noticed that this state of
`right-going' electrons corresponds precisely to the `maximum density
droplet' states which occur in the quantum Hall effect.) $n_{-qR}$
annihilates this state because there are no right-electron states with a
smaller total momentum than $|\Psi_0\rangle$. On the other hand for
$q = M 2 \pi / L$, $n_{qR}|\Psi_0\rangle$ yields a sum of $M$ terms in
which single-particle hole excitations have been formed in
$|\Psi_0\rangle$. For example, if we represent occupied states by solid
circles and unoccupied states by open circles, as in
Fig.~(\ref{fig:fermioncounting}), for $M=2$ we have
\begin{eqnarray}
n_{qR}\; |\Psi_{0}\rangle &=& |\ldots\bullet\;\bullet\;\circ\;\bullet |
\bullet\;\circ\;\circ\ldots\rangle\nonumber\\
 && + |\ldots\bullet\;\bullet\;\bullet\;\circ |
\circ\;\bullet\;\circ\ldots\rangle.
\label{eq:nov12e}
\end{eqnarray}
Each of the $M$ terms produced by $n_{qR}|\Psi_0\rangle$ is mapped back to
$| \Psi_0\rangle$ by $n_{-qR}$. Therefore
$n_{qR} n_{-qR} |\Psi_0\rangle = 0 $ whereas
$n_{-qR} n_{qR} |\Psi_0\rangle = M |\Psi_0 \rangle$. The general form of
the commutation relation is readily established by a little careful
algebra:\cite{mahanll}
\begin{equation}
[n_{-q'R},n_{qR}] = \frac{qL}{2\pi}\; \delta_{q,q'}.
\label{eq:nov12f}
\end{equation}
This holds as long as we truncate the Hilbert space to states with a
fixed number of right-going electrons and assume that states far from the
Fermi edge are always occupied.

We can define creation and annihilation operators for density wave
excitations of right-going electrons. For $q > 0$
\begin{eqnarray}
a_{q} &=& \sqrt{\frac{2\pi}{qL}}\; n_{-qR}\\
a_{q}^{\dagger} &=& \sqrt{\frac{2\pi}{qL}}\; n_{qR}
\end{eqnarray}
With these definitions Eq.~(\ref{eq:nov12f}) yields
\begin{equation}
[a_{q'},a^{\dagger}_q] = \delta_{q,q'}
\label{eq:nov27a}
\end{equation}
so that the density waves satisfy bosonic commutation relations.
Also note that
\begin{eqnarray}
\mbox{}[ \hat M,a_q] &=& - \frac{qL}{2 \pi} a_q \\
\mbox{}[ \hat M,a^{\dagger}_q ] &=& \frac{qL}{2 \pi} a^{\dagger}_q
\label{eq:nov27b}
\end{eqnarray}
where $\hat M$ is the total angular momentum operator.
The contribution to the Hamiltonian from right-going electrons is
therefore
\begin{equation}
H_{R} = \sum_{q>0} \hbar\; vq\; a_{q}^{\dagger} a_{q}^{\phantom{\dagger}}
\label{eq:hamiltonian}
\end{equation}
where
\begin{equation}
v = \frac{\alpha_{RR}}{2\pi\hbar} = \frac{1}{2\pi L\hbar}\;
\frac{d^{2}E_{0}}{dn_{R}^{2}} = \frac{1}{2\pi\hbar}\;
\frac{d\mu_{R}}{dn_{R}}
\label{eq:velocity}
\end{equation}
At low-energies the system is equivalent to a system of one-dimensional
phonons traveling to the right with velocity $v$. In the limit of
non-interacting electrons
\begin{equation}
v = \frac{\hbar k_{F}}{m^{\ast}} \equiv v_{F}
\end{equation}
as expected.

Without interactions between left and right-moving electrons a Luttinger
liquid is quite trivial. In particular the ground state
($|\Psi_0\rangle$) is a single-Slater determinant with a sharp Fermi
edge. For one-dimensional electron gas systems the interesting
physics\cite{mahanll} occurs only when left and right-moving electrons
are allowed to interact. Most notably, arbitrarily weak interactions
destroy the sharp Fermi edge which is the hallmark of Fermi liquids and
which survives interactions in higher dimensions. In the case of quantum
Hall edges, however, the above restriction to electrons moving in
only one direction is not a temporary pedagogical device.
The model with only right
moving electrons discussed above can be taken over {\em mutatis mutandis}
as a model of the edge excitations for an electron system with $\nu =1$.
The role played by the one-dimensional electron density
is taken over by the integral of
the two-dimensional electron density along a line perpendicular to the
edge. Results discussed in earlier sections can be discussed instead in
the language of chiral Luttinger liquids. The set of boson states are
the states of the chiral phonon system which has modes with only one sign
of momentum and velocity.

For $\nu = 1$ the analysis applies whether or not the electrons interact.
We now turn our attention to a discussion of the fractional case. Do all
steps of the above discussion generalize?  We can argue that if
we are interested only in low-energy long-wavelength excitations, the
energy can be expressed in the form
\begin{equation}
E = E_{0} + \frac{\alpha}{2L} \sum_{q\neq 0} n_{-q} n_{q}.
\end{equation}
As we comment later, this expression can fail at the edge of
fractional quantum Hall systems although it is appropriate for
$\nu = 1/m$.  What about the commutator?
There is an important difference in the line
of argument in this case, since single-particle states far from the edge
of the system are not certain to be occupied. Instead the average
occupation number is $\nu =1/m$ and there are large quantum fluctuations
in the local configuration of the system even in the interior. However,
we know\cite{macdedge} from the discussion in terms of many-body
wavefunctions in the previous section that the low-energy excitations at
$\nu =1/m$ {\em can} be described as the excitations of a boson system,
exactly like those at $\nu =1$,
which suggests that something like Eq.~(\ref{eq:nov12f})
must still be satisfied when the Hilbert space is projected to low
energies. If we replace the commutator by its expectation value in the
ground state we obtain
\begin{equation}
[n_{-q'},n_{q}] = \nu \cdot \frac{q L}{2\pi}\; \delta_{q,q'}
\label{eq:commutator}
\end{equation}
which differs from Eq.~(\ref{eq:nov12f}) only through the
factor $\nu$. It seems clear for the case of $\nu = 1/m$ this replacement
can be justified on the grounds that the the interior is essentially
frozen (but in this case {\em not} simply by the Pauli exclusion
principle) at excitation energies smaller than the gap for bulk
excitations. What we need to show is that Eq.~(\ref{eq:commutator})
applies as an operator identity in the entire low-energy portion of
the Hilbert space.  Below, however, we follow a different line of argument.

Appealing to the microscopic analysis in terms of many-body wavefunctions
we know that the excitation spectrum
for $\nu =1/m$ is equivalent to that of a
system of bosons. We conjecture that the commutator $[n_{-q'},n_{q}] =
\propto q \delta_{q,q'}$. To determine the constant of proportionality
we will require that the rate of change of the equilibrium edge current with
chemical potential be
$ e \nu / h$. From the edge state picture of the quantum Hall effect
discussed in Section II, it is clear that this is equivalent to requiring
the Hall conductivity to be quantized at
$\nu e^2 /h$. Since our theory will yield a set of phonon modes which
travel with a common velocity $v$ it is clear that the change in
equilibrium edge current is related to the change in equilibrium density
by
\begin{equation}
\delta I = ev\delta n.
\end{equation}
When the chemical potential for the single edge system is shifted
slightly from its reference value (which we chose to be zero) the grand
potential is given by
\begin{equation}
E[n] = E_{0} + \mu\delta n + \alpha\frac{(\delta n)^{2}}{2}
\label{eq:nov13a}
\end{equation}
Minimizing with respect to $\delta n$ we find that
\begin{equation}
\delta n = \frac{\delta \mu}{\alpha}
\label{eq:nov13b}
\end{equation}
so that
\begin{equation}
\frac{\delta I}{\delta\mu}  = \frac{ e v }{\alpha}
\label{eq:nov13c}
\end{equation}
In order for this to be consistent with the quantum Hall effect ($ \delta
I = (e \nu /h) \delta \mu $) our theory must yield a edge phonon
velocity given by
\begin{equation}
 v  = \frac{\alpha}{h} \cdot \nu.
\label{eq:nov13d}
\end{equation}
The extra factor of $\nu$ appearing in this equation compared to
Eq.~(\ref{eq:velocity}) requires the same factor of $\nu$ to appear in
Eq.~(\ref{eq:commutator}). We discuss below the qualitative changes in the
physics\cite{wen,wenreview} of fractional edge states
which are implied by this outwardly innocent numerical factor.

It is worth remarking that the line of argument leading to this specific
chiral Luttinger liquid theory of the fractional quantum Hall effect is
not completely rigorous. In fact we know that this simplest possible
theory with a single branch of chiral bosons does not apply for all
filling factors,\cite{macdedge,wen,macdj} even though (nearly) all steps
in the argument are superficially completely general. The reader is
encouraged to think seriously about what could go wrong with our
arguments. Certainly the possibility of adiabatically connecting all
low-energy states with corresponding states of the non-interacting
electron system, available for one-dimensional electron gases and for
quantum Hall systems at integer filling factors but not at fractional
filling factors, adds confidence when it is available. In our view, the
microscopic many-particle wavefunction approach which establishes a
one-to-one mapping between integer and fractional edge excitations (for
$\nu =1/m$!) is an important part of the theoretical underpinning of the
Luttinger liquid model of fractional Hall edges.  Once
we know that the edge excitations map to those of a
chiral boson gas and that the
fractional quantum Hall effect occurs, it appears that no freedom is left
in the construction of a low-energy long-wavelength effective theory. The
reader is reminded however, of the smooth edge regime, where in our view
both the many-particle wavefunction mapping and the chiral Luttinger
liquid theory of the edge are likely to fail.

An important aspect of Luttinger liquid theory is the expression for
electron field operators in terms of bosons.\cite{mahanll} This
relationship is established by requiring the exact identity
\begin{equation}
[\rho (x),\hat{\psi}^{\dagger}(x')] = \delta (x-x')\;
\hat{\psi}^{\dagger}(x')\;
\label{eq:nov13a1}
\end{equation}
to be reproduced by the effective low-energy theory. This equation simply
requires the electron charge density to increase by the required amount
when an electron is added to the system. The electron creation operator
should also be consistent with Fermi statistics for the electrons:
\begin{equation}
\left\lbrace\psi^{\dagger}(x),\psi^{\dagger}(x')\right\rbrace =0 .
\label{eq:nov13a2}
\end{equation}
In order to satisfy Eq.~(\ref{eq:nov13a1}), the field operator must be
given by
\begin{equation}
\hat{\psi}^{\dagger}(x) = c e^{i\nu^{-1}\phi (x)}
\label{eq:nov13a3}
\end{equation}
where $d \phi(x) / dx = n(x)$ and $c$ is a constant which cannot be
determined by the theory. The factor of $\nu^{-1}$ in the argument of
the exponential of Eq.~(\ref{eq:nov13a3}) is required because of the
factor of $\nu$ in the commutator of density Fourier components which
in turn was required to make the theory consistent with the fractional quantum
Hall effect.  When the exponential is expanded the $k-th$ order
terms generate states with total boson occupation number
$k$ and are multiplied in the fractional case by the
factor $\nu^{-k}$; multi-phonon terms are increased in importance. It is
worth remarking\cite{wen} that the anticommutation relation between
fermion creation operators in the effective theory is satisfied only when
$\nu^{-1}$ is an odd integer. This provides an indication, independent of
microscopic considerations, that the simplest single-branch chiral boson
effective Hamiltonian can be correct only when
$\nu =1/m$ for odd $m$. Wen\cite{wenreview} has surveyed, using this
criterion, the multi-branch generalization of the simplest effective
Hamiltonian theory which are possible at any given rational filling
factor. His conclusions are consistent with arguments\cite{macdedge}
based on the microscopic theory of the fractional quantum Hall effect.

Eq.~(\ref{eq:nov13a3}) has been carefully checked
numerically\cite{palacios} and appears to be correct. The $\nu^{-1}$
factor leads to predictions of qualitative changes in a number of
properties of fractional edges. The quantity which is most directly
altered is the tunneling density-of-states. Consider, for example, the
state created when an electron, localized on a magnetic length scale, is
added to the ground state at the edge of a $N-$ electron
system with $\nu =1/m$:
\begin{eqnarray}
\hat{\psi}^{\dagger}(0) |\Psi_{0}\rangle &\sim& \exp{\left(
-\sum_{n>0} \frac{a_{n}^{\dagger}}{\sqrt{n \nu}}\right)}
|\psi_{0}\rangle\nonumber\\
 &=& 1 + \frac{1\mbox{ phonon term}}{\nu^{1/2}} + \frac{2\mbox{ phonon
terms}}{\nu } + \ldots.
\end{eqnarray}
The tunneling density states is given by a sum over the ground and
excited states of the $N+1$ particle system:
\begin{equation}
A(\epsilon) = \sum_n \delta(E_n - E_0 - \epsilon)
| \langle \Psi_n | \psi^{\dagger}(0) \Psi_0 | \rangle |^2
\label{eq:sfun}
\end{equation}
Because of the increased weighting of multiphonon states, which become
more numerous at energies farther from the chemical potential,
the spectral function is larger at larger $\epsilon -
\mu$ in the fractional case. An explicit calculation\cite{wen,wenreview}
yields a spectral function which grows like
$(\epsilon-\mu)^{{\nu}^{-1}-1}$. It is intuitively clear that the
spectral function should be small at low-energies in the fractional case
since the added electron will not share the very specific correlations
common to all the low-energy states. It is amazing that by simply
requiring the low-energy theory to be consistent with the fractional
quantum Hall effect we get a very specific prediction for the way in
which this qualitative notion is manifested in the tunneling density of
states.

\section{Endnote}

There is, as always, a lot more which could be said. However other
duties, including even ones for which I am paid, are insisting that I
must stop here. These notes have focused on the
microscopic origins, and also possibly (in the `smooth edge' case)
the limitations of Luttinger liquid theories for quantum Hall edges.
A different and at least equally interesting article could be written on
applications of Luttinger liquid models in the fractional quantum
Hall regime.  Among these it appears that
those which describe\cite{twoedgecaveat} tunneling\cite{restun}
between quantum Hall edge systems due to disorder are the most promising
for experimental tests. Indeed, initial experiments\cite{milliken} appear
to confirm theoretical predictions. It is likely that more experimental
work will be added to the literature soon. The important question of the
role of long-range interactions in Luttinger liquid theories, which we
have for the most part dodged here, is also beginning
to be addressed.\cite{longrange}

These informal notes are intended to be widely accessible. I hope that
they will be of some use to people who are expert on either theoretical or
experimental aspects of the quantum Hall effect but have not been
following theories of quantum Hall edges.  Likewise I hope
that they can be useful to experts on the one-dimensional electron gas who
have not been following the quantum Hall effect.  Comments, critical or
complimentary, are welcome. The ideas here have been shaped by
discussions with members of the condensed matter theory group at Indiana
University, especially S.M. Girvin, R. Haussmann, S. Mitra,
K. Moon, J.J. Palacios, D. Pfannkuche,
E. Sorensen, K. Tevosyan, K. Yang, and U. Z\"{u}licke.
Discussions with L. Brey, M. Fisher, M. Johnson,
C. Kane, L. Martin, J. Oaknin,
C. Tejedor, S.R.-E. Yang and X.-G. Wen are also gratefully
acknowledged.  The responsibility for surviving misapprehensions
rests with me. This work was supported by the National Science Foundation
under grant DMR-9416906.

\begin{figure}
\caption{A large but finite two-dimensional electron gas. In panel (a) the
chemical potential lies in a gap and the only low-energy excitations are
localized at the edge of the system. In panel (b) the chemical potential
lies in a mobility gap so that there are low-energy excitations in the
bulk but they are localized away from the edge. In panel (c) a net current
is carried from source to drain by having local equilibria at different
chemical potentials on upper and lower edges.}
\label{fig:incomp}
\end{figure}

\begin{figure}
\caption{Schematic spectrum for non-interacting electrons confined to a
circular disk in a strong magnetic field. In the limit of large disks
the dependence of the energy on $m$ can usually be considered to be
continuous. The situation depicted has Landau level filling factor $\nu
=2$ in the bulk of the system.}
\label{fig:nonintel}
\end{figure}

\begin{figure}
\caption{Non-interacting many electron eigenstates for small excess
angular momentum $M$ specified by occupation numbers for the
single-particle states with energies near the chemical potential $\mu$.
The vertical bars separate single-particle states with $\epsilon_m < \mu$
from those with $\epsilon_m > \mu$. A solid circle indicates that $n_m=1$
in both the ground state and in the particular excited state; a shaded
circle indicates that
$n_m = 1$ in the particular excited state but not in the ground state; an
empty circle indicates that $n_m =0$.}
\label{fig:fermioncounting}
\end{figure}

\begin{figure}
\caption{Schematic illustration of the Thomas-Fermi theory and
Hartree-Fock theory pictures of the edge of a quantum Hall system with
Landau level filling factor $\nu =2$ in the bulk. The Thomas-Fermi theory
gives an approximation for the charge density profile across the edge.
Regions where the charge density varies, indicated by dashed lines,
correspond to constant values for $V_{\rm ext} + V_H $ and are known as
compressible strips. The edge excitations occur in the compressible
strips. Hartree-Fock theory produces approximate values for the
quasiparticle energies as a function of angular momentum in the edge
region. When the electron density at the edge varies gradually on an
atomic length scale the Hartree-Fock eigenvalues will have a weak
dependence on angular momentum where they cross the Fermi level.}
\label{fig:thfehf}
\end{figure}

% tables follow here
\begin{table}
\caption{Quantum occupation numbers in boson and fermion descriptions for
edge excitations with small excess angular momentum $M$. $g_M$ is the
number of states with excess angular momentum $M$. The fermion occupation
numbers are relative to the maximum density droplet state. Only
non-zero values are listed for both fermion and boson descriptions.
$L=N-1$ is the highest angular momentum which is occupied in the maximum
density droplet state.}
\label{table1}
\begin{tabular}{llll}
M & $g_M$ & \underline{Fermion Description} & \underline{Boson Description}\\
1 & 1 & $ n_{L+1} = 1, n_{L} = -1$ & $n_{1} = 1$\\
2 & 2 & $ n_{L+2} = 1, n_{L} = -1; n_{L+1} = 1, n_{L-1} = -1$ &
$n_{2} = 1; n_{1} = 2$\\
3 & 3 & $ n_{L+3}=1,n_L=-1;n_{L+2}=1,n_{L-1}=-1$
& $n_{3} = 1; n_{2} = 1, n_{1} = 1;$\\
 & & $ n_{L+1}=1 n_{L-2} =-1$
& $n_{1} = 3$\\
4 & 5 & $ n_{L+4}=1, n_{L}=-1; n_{L+3}=1,n_{L-1}=-1
$& $n_{4}=1;n_{3}=1,n_{1}=1;n_{2}=2$ \\
 & & $ n_{L+2}=1,n_{L-2}=-1; n_{L+1}=1,n_{L-3}=-1$ &
$n_{2}=1,n_{1}=2; n_{1}=4$ \\
 & & $ n_{L+2}=1,n_{L+1}=1,n_{L}=-1,n_{L-1}=-1$& \\
\end{tabular}
\end{table}


\begin{references}
\bibitem{negative} For negatively charged particles the angular momenta in
the lowest Landau level are negative for a magnetic field in the
$\hat z$ direction. We choose a magnetic field in the $- \hat z$
direction to have the convenience of positive angular momenta in the
lowest Landau level.
\bibitem{caveat} The antisymmetry requirement generally requires
$P[z]$ must be an odd function of $z_i-z_j$ for any $i$ and $j$ so that
$P[z]$ contains only odd terms in a Taylor series expansion of its
dependence on this difference coordinate. The coefficient of the
$(z_i-z_j)^1$ term must vanish if the pair never has relative angular
momentum equal to one so that the leading term is $(z_i-z_j)^3$. It then
follows from analyticity that each difference coordinate to the third
power is a factor of $P[z]$.
\end{references}
\end{document}